\documentclass[intlimits,twoside,a4paper]{article}

\usepackage[cp1251]{inputenc}
\usepackage[eqsecnum]{cmpj3}

\issue{2021}{24}{4}{43704}
\doinumber{10.5488/CMP.24.43704}
\title[DFT based investigation of solid-state In$_{1-x}$Ti$_{x}$Sb solutions]%
{DFT based investigation of the structural, magnetic, electronic, and half-metallic properties of solid In$_{1-x}$Ti$_{x}$Sb solutions}%

\author[S. Amrani, M. Berber, M. Mebrek]{S. Amrani\orcid{0000-0002-3484-6480}\refaddr{label1}, M. Berber\orcid{0000-0003-1285-3070}\refaddr{label2,label3} \thanks{Corresponding author: \email{berbermohamed@yahoo.fr}.}, M. Mebrek\orcid{0000-0002-4116-1332}\refaddr{label2,label3}}
\addresses{
\addr{label1} Laboratory of Physico-Chemical Studies, University of Saida, 20000 Saida, Algeria
\addr{label2} Advanced Materials and Instrumentation Laboratory, Nour El-Bachir El-Bayadh University center, Algeria
\addr{label3} Nour El-Bachir El-Bayadh University center, 32000 El Bayadh, Algeria
}


\date{Received Septmeber 02, 2021, in final form October 30, 2021}

\begin{document}

\maketitle

\begin{abstract}

With the intention to reveal the effect of the substitution, Ti-doped InSb alloy, we accomplished a first-principles prediction within the FPLAPW+lo method. We used GGA-PBEsol scheme attached with the improved TB-mBJ approach to predict structural, electronic, and magnetic properties of In$_{1-x}$Ti$_{x}$Sb with concentration $x = 0$, $0.125$, $0.25$, $0.50$, $0.75$, $0.875$, and $1$. Our lattice parameters are found in favorable agreement with the available theoretical and experimental data. The calculation shows that all structures are energetically stable. The substitutional doping transforms the ionic character of the InSb compound in half-metallic ferromagnetic comportment for concentration $x = 0$, $0.125$, $0.25$, and $0.50$, with a spin polarization of $100$\% at the Fermi level, and metallic nature for In$_{0.25}$Ti$_{0.75}$Sb and In$_{0.125}$Ti$_{0.875}$Sb. The total magnetic moments are also estimated at about 1~$\mu_{\text{B}}$.  In$_{0.875}$Ti$_{0.125}$Sb,  In$_{0.75}$Ti$_{0.25}$Sb, and In$_{0.50}$Ti$_{0.50}$Sb have half-metallic ferromagnets comportment and they can be upcoming applicants for spintronics applications.

\printkeywords
%
\end{abstract}

\section{Introduction}


The possibility to dope the semiconductors with magnetic elements has created a new category of materials called diluted magnetic semiconductors (DMS). One of the peculiarities of DMS is the appearance of localized magnetic moments generated by electron-hole coupling~\cite{volkov1978interaction}. This leads to interesting properties, for example, increasing the $g$ factor of the charge carriers or new excitements, such as magnetic polarons~\cite{yakovlev2010magnetic}. However, ferromagnetism is rarely observed in semiconductors due to the low density of carriers and because of the predominance of superexchange between local magnetic moments. In most cases, magnetic semiconductors have low Curie temperatures. Interest in DMSs was recently aroused by theoretical and experimental work on  semiconductors~\cite{laroussi2019effect,laroussi2019first,berber2017investigation,berber2018first,bahloul2019electronic,boutaleb2021study,bouziani2020high,khan2020theoretical,marquina2020theoretical,
monir2020density,korichi2020ferromagnetism,hamidane2020ab,maftouh2021first,
kumar2021structural,abo2021structure,yehia2021structural,zhao2021magnetic}. DMS based on the transition element doped and on semiconductors has two important characteristics. The first one, the transition concentration of the elements can be up to ten percent (using molecular beam epitaxy techniques, MBE)~\cite{cho1975molecular}. The second one, the concentration of carriers can be controlled in a wide range and permits to choose between $n$-type and $p$-type doping. These two aspects offer the possibility of varying the magnetic behavior by changing the concentration of the carriers.
Indium Antimonide is one of the best frequently exploited semiconductors, which is used in Hall effect sensors, thermoelectric generators, infrared detectors, high-speed photodetectors, and transistors~\cite{sandhu2004high,nilsson2009giant,kuo2013high,minamoto1962insb,maffitt1964insb,hattori1982electrical,kimukin2003insb, ashley2007heterogeneous,singh2017study,fawcett1969negative,ashley1995uncooled}. This extensive utilization follows from its distinct properties, such as a narrow band gap, small effective electron mass, and high electron mobility. Doping different materials into the InSb structure fascinates several scientists. Consequently, the category of dopant strongly influences InSb characteristics and should be precisely chosen according to suitable properties and future applications. The optimization of pulse electrodeposition method for the synthesis of Te-doped InSb films was performed by Rajska et al.~\cite{rajska2021optimization}. They show that the bandgap values affirm an impact of a dopant content on the optical properties of thin Te-doped InSb films. The maximum figure of merit (ZT) was found to be about 0.6 and 673~K for Te-doped InSb~\cite{yamaguchi2005thermoelectric}. They conclude that this material is a very promising candidate for thermoelectric devices. The transport properties of Sn-doped InSb, were experimentally studied by Okamoto and Shibasaki~\cite{okamoto2003transport} and confirm its use in Hall sensors applications. InSb doped with magnetic elements for spintronics applications has been the patience of several researchers. Theoretical and experimental studies have been carried out to describe the magnetic phenomenon and the half-metallicity comportment. Mn-doped InSb was studied by Novotortsev et al.~\cite{kochura2018magnetotransport}, (Fe,~Mn,~and~Cr) doped InSb~\cite{you2020microscopic, mir2020exploration,berber2020spinelectronic,pappert2006magnetization,ohno1996ga,sato2000material,
ruderman1954indirect,yosida1957magnetic}.
The aim of our study is to predict structural, electronic, and magnetic properties of Ti-doped InSb. 

\begin{figure}[!h]
	\centerline{\includegraphics[width=0.65\textwidth]{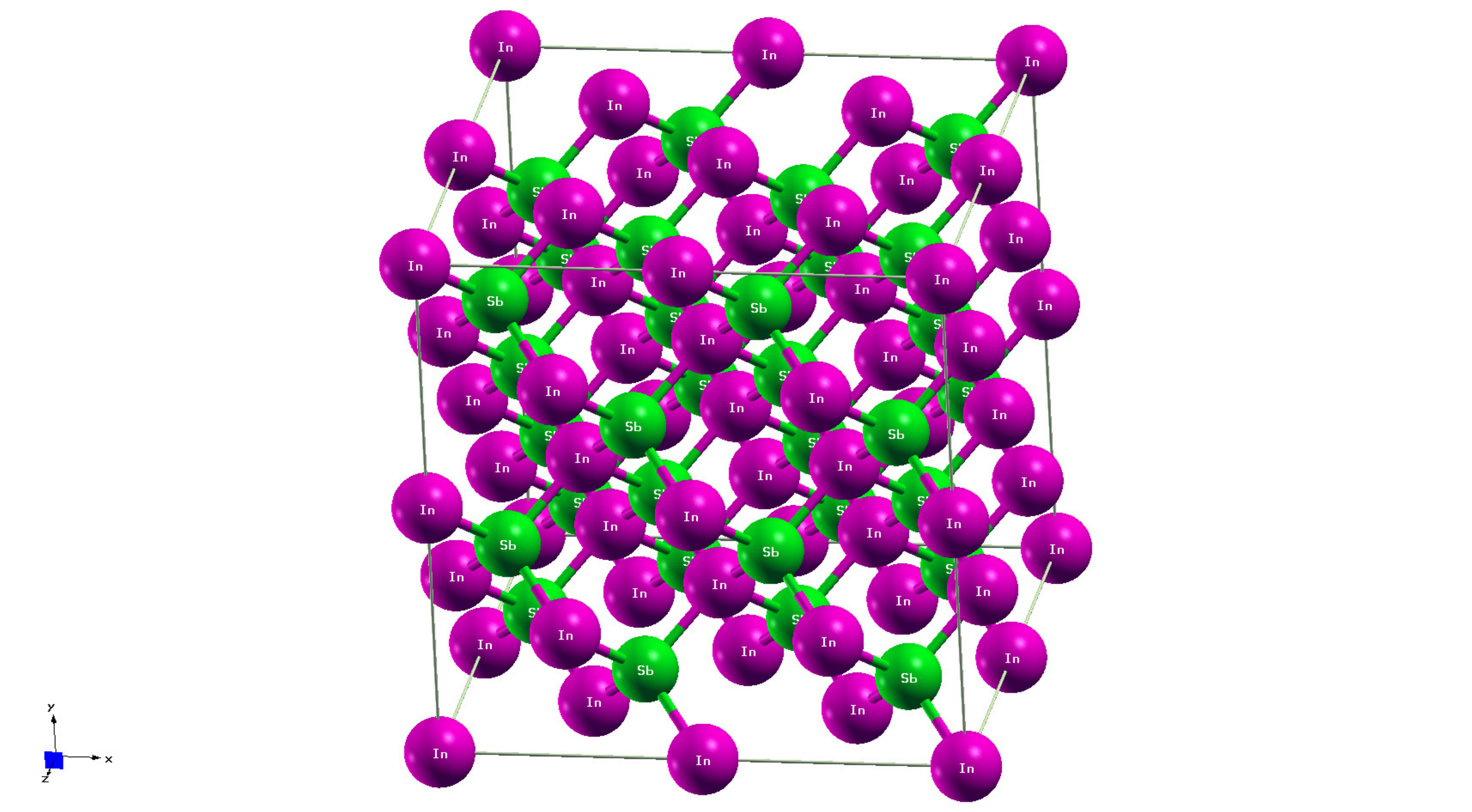}}
	\caption{(Colour online) Crystal supercell structure of InSb.} \label{fig1}
\end{figure}

\section{Computational method}

With the intention of calculating appropriate properties of Ti-doped InSb, we performed calculations within the full-potential linearized augmented plane wave (FP-LAPW) method~\cite{mun2013first,khalil2014magnetic,berri2015ab,sjostedt2000alternative} based on the density functional theory (DFT)~\cite{sjostedt2000alternative,blaha2001wien2k}  implemented in WIEN2k code~\cite{blaha2001wien2k}. We employed Perdew--Burke--Ernzerhof GGA-PBEsol approximation~\cite{perdew1996generalized}, combined with Tran--Blaha modified Becke--Johnson TB-mBJ approximation~\cite{koller2011merits,koller2012improving,tran2009accurate}.In our calculation, we used the following parameters:

\begin{itemize}
\item 	$K_{mesh}$ 200 k-points in Brillion zone (BZ).
\item 	The muffin-tin radii for Ti, In, and Sb are chosen as 1.65, 1.85, and 1.96~a.u., respectively.
\item  $RMT \cdot K_{\rm{max}}=7$, where $RMT$ is the smallest value of muffin-tin sphere radius, and $K_{\rm{max}}$ indicates the largest vector in-plane wave expansion.
\item 	$l_{\rm{max}}= 6$, where $l_{\rm{max}}$ is the wave function inside the muffin-tin spheres.
\item 	$G_{\rm{max}} = 14$, where $G_{\rm{max}}$ is the magnitude largest vector in the charge density Fourier expansion.
\item 	The cut-off energy selected as $-6$~Ry (the separation of valence and core states ).
\item	The self-consistent calculations are converged when the condition is set to 0.0001~Ry.
\item 	Employing configuration $2 \times 2 \times 2$, a supercell of In$_8$Sb$_8$ with (Zincblende, F4 3m, space group N:216) was constructed, which, consists of 16~atoms in which 8~atoms belong to Indium and 8~atoms belong to Antimonide. The doping of Titanium cation was done on an In$_{1-x}$Ti$_{x}$Sb structure. In$_{0.875}$Ti$_{0.125}$Sb, In$_{0.75}$Ti$_{0.25}$Sb, In$_{0.50}$Ti$_{0.50}$Sb, In$_{0.25}$Ti$_{0.75}$Sb, and 
In$_{0.125}$Ti$_{0.875}$Sb structures were created by Ti atom on substitution sites with Indium. The relaxation maintained the F4 3m symmetry for the structure investigated.
\end{itemize}

\begin{table}[htb]
	\caption{Calculated lattice constant ($a$), bulk modulus ($B$), and its pressure derivative ($B'$) for for In$_{1-x}$Ti$_{x}$Sb at concentrations $x = 0$, $0.125$, $0.25$, $0.50$, $0.75$, $0.875$ and $1$.}
	\label{tbl-smp1}
	\vspace{2ex}
	\begin{center}
		\vspace{2ex}
		\begin{tabular}{|c|c|c|c|c|c|c|}
			\hline 
			\small Compound &\small $a$  (\textup{\AA}) & \small $B$ (GPa) &\small $B'$ &  \small $ E_{\rm {Form}}$ &\small $ E_{\rm{coh}}$ & \small Method \\ 
			\hline 
			\small This Work &  &  & &  &  & \small GGA-WC \\ 
			\hline 
			\small InSb & \small 6.5554 & \small 42.4612 & \small 4.0525 &  &  &  \\ 
			\hline 
			\small In$_{0.875}$Ti$_{0.125}$Sb & \small 6.4876, 6.5069~\cite{vegard1921formation}  & \small 39.7953 & \small 5.0000 & \small $-0.341$ & \small 0.941 & \\ 
			\hline 
			\small In$_{0.75}$Ti$_{0.25}$Sb& \small 6.4455, 6.4585~\cite{vegard1921formation} & \small 54.2042 & \small 5.0000 & \small $-0.522$& \small 0.852 &  \\ 
			\hline 
			\small In$_{0.50}$Ti$_{0.50}$Sb & \small 6.3878, 6.3616~\cite{vegard1921formation}   & \small 52.0305& \small 6.7370 & \small $-0.435$& \small 0.902 &  \\ 
			\hline 
			\small In$_{0.25}$Ti$_{0.75}$Sb& \small 6.3087, 6.2648~\cite{vegard1921formation} & \small 53.8414 &\small 2.9134& \small $-0.476$& \small 0.871 &  \\ 
			\hline 
			\small In$_{0.125}$Ti$_{0.875}$Sb & \small 6.2315, 6.2164~\cite{vegard1921formation} & \small 62.1805 & \small 4.7886 & \small $-0.393$ & \small 0.932 &  \\ 
			\hline 
			\small TiSb &\small 6.1697 & \small 58.0755 & \small 4.3008&  &  &  \\ 
			\hline 
			\small 	Other calculations & &  &  &  & &  \\ 
			\hline 
			\small InSb  & \small 6.464, 6.36~\cite{ashley2007heterogeneous}&  &  &  &  &  \\ 
			\hline 
			&\small  6.479~\cite{singh2017study}&  &  &  &  &  \\ 
			\hline 
			&\small  6.643~\cite{fawcett1969negative}&  &  &  &  &  \\ 
			\hline 
			&\small  6.629~\cite{ashley1995uncooled}&\small  37.368~\cite{ashley1995uncooled}& \small  4.452~\cite{ashley1995uncooled}& &  &\small GGA-PBE  \\ 
			\hline 
			&\small  6.47937~\cite{rajska2021optimization}, 6.4791~\cite{breivik2013temperature}&  &  &  &  &\small Experimental  \\ 
			\hline 
			&\small  6.49~\cite{okamoto2003transport}, 6.47937~\cite{adachi1999indium}&  & & &  &\small Experimental \\ 
			\hline 
		\end{tabular} 
		\renewcommand{\arraystretch}{1}
	\end{center}
\end{table}

\begin{figure}[!h]
	\centerline{\includegraphics[width=0.90\textwidth]{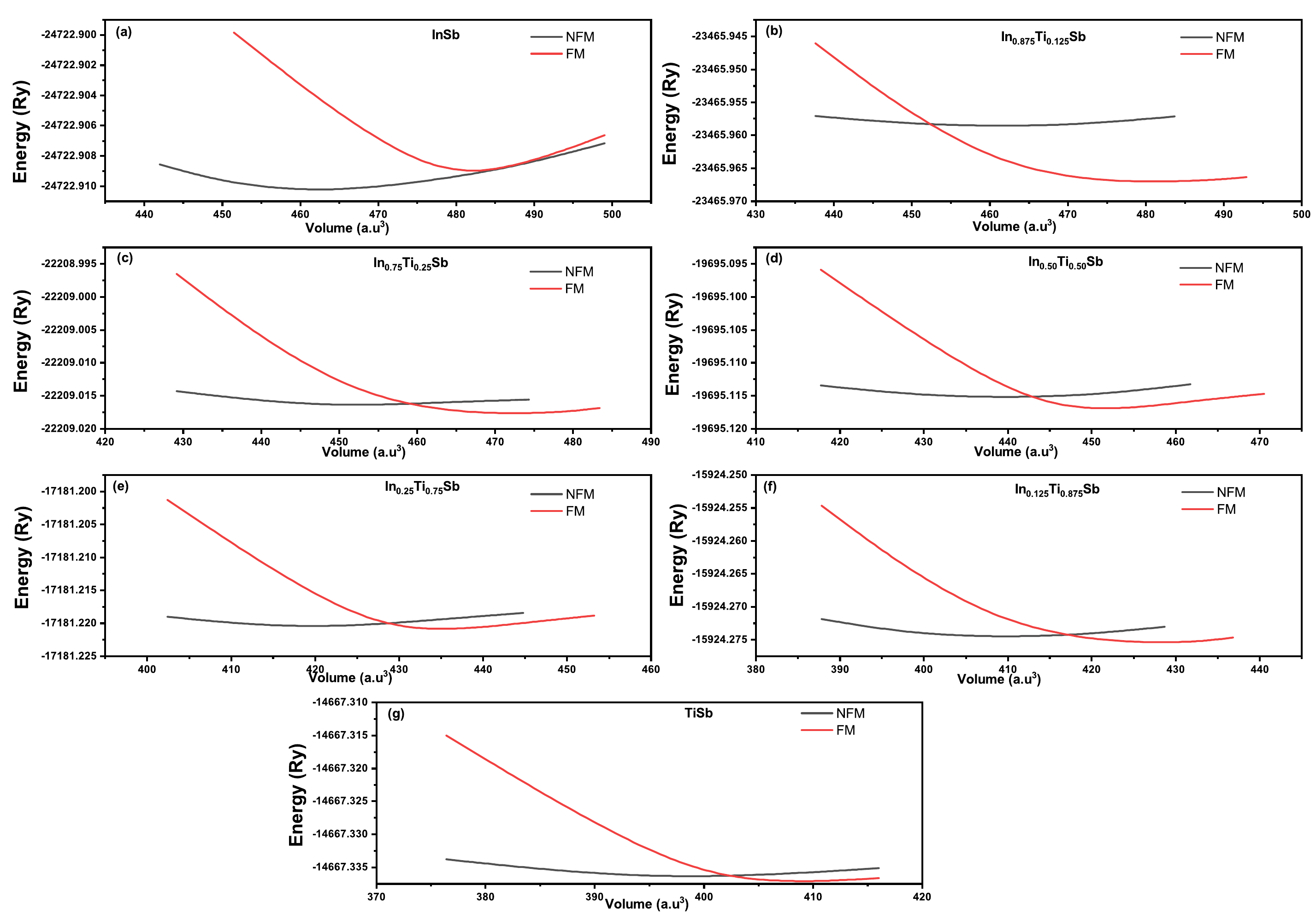}}
	\caption{(Colour online) Energy unit cell versus volume with FM and NFM phase of (a)~InSb, (b)~In$_{0.875}$Ti$_{0.125}$Sb, (c)~In$_{0.75}$Ti$_{0.25}$Sb, (d)~In$_{0.50}$Ti$_{0.50}$Sb, (e)~In$_{0.25}$Ti$_{0.75}$Sb, (f)~In$_{0.125}$Ti$_{0.875}$Sb, and (g)~TiSb.} \label{fig2}
\end{figure}

\section{Results and discussion}

\subsection{Structural properties}

Indium Antimonide (InSb) crystallizes in the zincblende structure (No.~216). Figure~\ref{fig1} exhibits the crystal supercell structure of InSb. Its conventional structure has two types of atoms, In and Sb, located at (0, 0, 0) and (0.25, 0.25, 0.25) positions, respectively~\cite{liu1951lattice,orton2009semiconductors}. The In$_{0.875}$Ti$_{0.125}$Sb, In$_{0.75}$Ti$_{0.25}$Sb, In$_{0.50}$Ti$_{0.50}$Sb, In$_{0.25}$Ti$_{0.75}$Sb, and 
In$_{0.125}$Ti$_{0.875}$Sb compounds are constructed by substitutional doping of one ($1/8=0.125$), two ($2/8=0.25$), four ($4/8=0.50$), six ($6/8=0.75$) and seven ($7/8=0.875$) atoms of Titanium (Ti) at Indium (In) sites, respectively.
 The structures of In$_{1-x}$Ti$_{x}$Sb with concentration $x = 0$~(InSb), 0.125, 0.25, 0.50, 0.75, 0.875, and 1 (TiSb) are completely optimized and relaxed. For both ferromagnetic (FM) and non-ferromagnetic (NFM) phases, the energy curves as a function of the volume of each structure (see figure~\ref{fig2}) are calculated and fitted by the suitableness of Birch--Murnaghan's equation of state~(EOS)~\cite{birch1947finite}. According to figure~\ref{fig2}, all the structures are stable in the FM phase except the binary InSb. In this respect, we estimated the structural properties including the equilibrium lattice constant ``$a$'', bulk modulus ``$B_0$'', and its derivative $B'$. The results are collected in table~\ref{tbl-smp1} accompanied by the available theoretical and experimental data. 
The thermodynamic stability and the probability of synthesizing these alloys~\cite{reunchan2012mechanism,liu2017electronic} are determined by the formation and cohesive energies. To confirm the phase stability of the In$_{0.875}$Ti$_{0.125}$Sb, In$_{0.75}$Ti$_{0.25}$Sb, In$_{0.50}$Ti$_{0.50}$Sb, In$_{0.25}$Ti$_{0.75}$Sb, and 
In$_{0.125}$Ti$_{0.875}$Sb  structures, we calculated the formation and cohesive energies using the following expressions~\cite{yosida1957magnetic, mun2013first}:

\begin{figure}[!htb]
	\centerline{\includegraphics[width=0.45\textwidth]{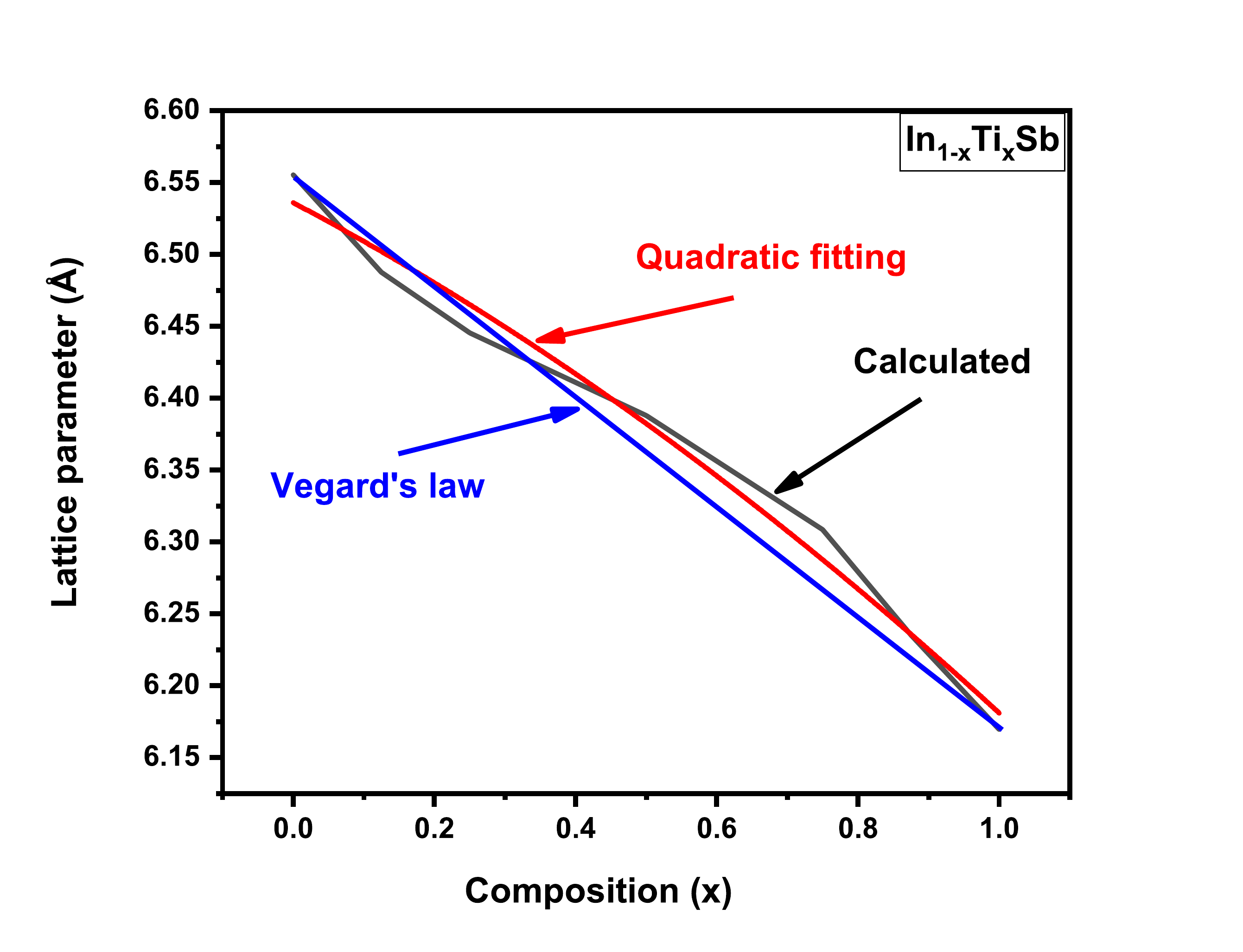}}
	\caption{(Colour online) The calculated lattice constant $a$ of In$_{1-x}$Ti$_{x}$Sb as a function of concentration~($x$).} \label{fig3}
\end{figure}
\begin{align}
E^{\rm {In}_{1-x}\rm {Ti}_{x}\rm {Sb}}_{\rm {form}}=E^{\rm {In}_{1-x}\rm {Ti}_{x}\rm {Sb}}_{\rm {\rm {total}}}-\left[(1-x)E^{\rm{In}}_{\rm{solid}}+xE^{\rm{Ti}}_{\rm{solid}}+8E^{\rm{Sb}}_{\rm{solid}}\right],
\label{energie de formation}
\end{align}
\begin{align}
E^{\rm {In}_{1-x}\rm {Ti}_{x}\rm {Sb}}_{\rm{coh}}=\left[(1-x)E^{\rm{In}}_{\rm{atom}}+xE^{\rm{Ti}}_{\rm{atom}}+8E^{\rm{Sb}}_{\rm{atom}}\right]-E^{\rm {In}_{1-x}\rm {Ti}_{x}\rm {Sb}}_{\rm {total}}.
\label{energie de cohesion}
\end{align}
Here $E^{\rm {In}_{1-x}\rm {Ti}_{x}\rm {Sb}}_{\rm {total}}$  is the total energy per atom of In$_{1-x}$Ti$_{x}$Sb and $E^{\rm{In}}_{\rm{solid}}$, $E^{\rm{Ti}}_{\rm{solid}}$, and $E^{\rm{Sb}}_{\rm{solid}}$ are the total energies per atom of In, Ti, and Sb solids, respectively. $E^{\rm{In}}_{\rm{atom}}$, $E^{\rm{Ti}}_{\rm{atom}}$, and $E^{\rm{Sb}}_{\rm{atom}}$  are the isolated atomic energies of the pure constituents of In, Ti, and Sb, respectively. Our computed values of formation and cohesive energies for In$_{0.875}$Ti$_{0.125}$Sb, In$_{0.75}$Ti$_{0.25}$Sb, In$_{0.50}$Ti$_{0.50}$Sb, In$_{0.25}$Ti$_{0.75}$Sb, and 
In$_{0.125}$Ti$_{0.875}$Sb compounds are collected in table~\ref{tbl-smp1}. Therefore, the negative values of the formation energy and the positive values of cohesive energy, imply that the ternary alloys In$_{1-x}$Ti$_{x}$Sb with concentration $x = 0$~(InSb), 0.125, 0.25, 0.50, 0.75, 0.875, and 1 (TiSb) are energetically and thermodynamically stable and can be synthesized under normal conditions.
The deviation of a lattice constant $a$ as a function of the composition~($x$) for In$_{1-x}$Ti$_{x}$Sb is shown in figure~\ref{fig3}. The linear relationship between the lattice constant and the dopant concentration $x$~\cite{vegard1921formation} is predicted by Vegard's law as follows:
\begin{align}
a(x)=6.5554-0.3875x.
\label{loi de vegard}
\end{align}

However, the quadratic relationship is widely used for both experimental and theoretical purposes.  We obtained the fitting equation for In$_{1-x}$Ti$_{x}$Sb:
\begin{align}
a(x)=6.53614-0.26049x-0.09602x^{2}.
\label{fitting}
\end{align}

\begin{figure}[htb]
\centerline{\includegraphics[width=0.65\textwidth]{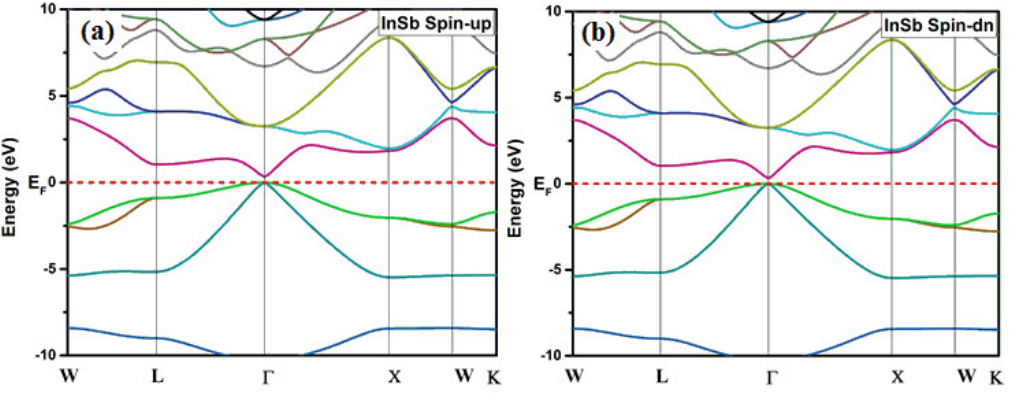}}
\caption{(Colour online) Spin-polarized band structure obtained with TB-mBJ for InSb: (a)~majority spin (up) and (b) minority spin (dn). The Fermi level is set to zero (horizontal dotted red line).} \label{fig4}
\end{figure}

\begin{figure}[htb]
\centerline{\includegraphics[width=0.65\textwidth]{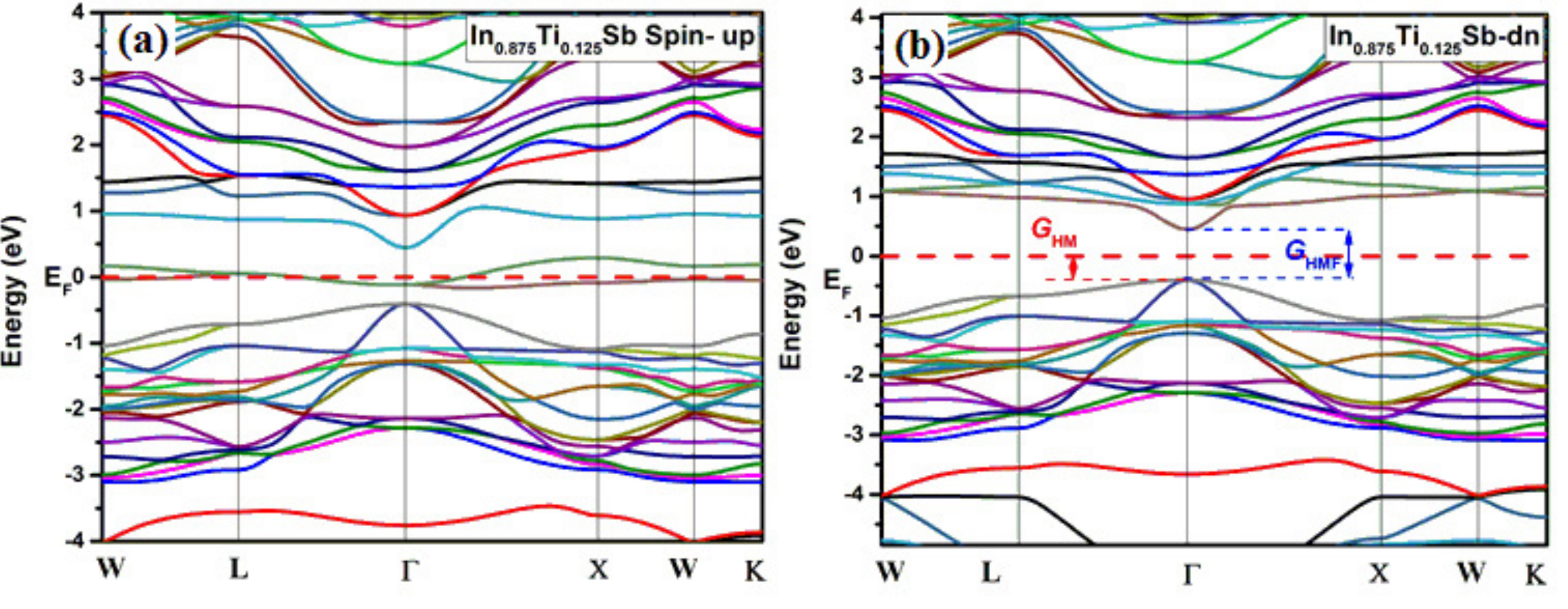}}
\caption{(Colour online) Spin-polarized band structure obtained with TB-mBJ for In$_{0.875}$Ti$_{0.125}$Sb: (a)~majority spin (up) and (b) minority spin (dn). The Fermi level is set to zero (horizontal dotted red line).} \label{fig5}
\end{figure}

\begin{figure}[htb]
\centerline{\includegraphics[width=0.65\textwidth]{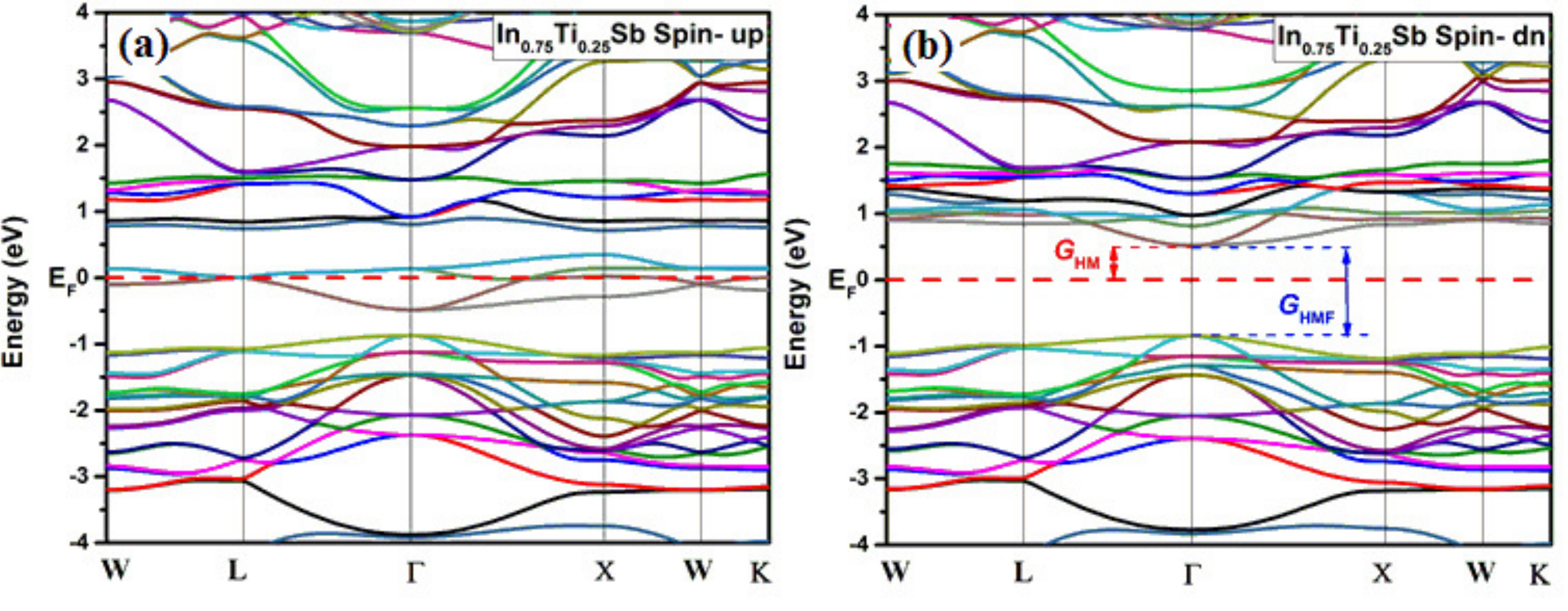}}
\caption{(Colour online) Spin-polarized band structure obtained with TB-mBJ for In$_{0.75}$Ti$_{0.25}$Sb: (a)~majority spin (up) and (b) minority spin (dn). The Fermi level is set to zero (horizontal dotted red line).} \label{fig6}
\end{figure}

\begin{figure}[htb]
\centerline{\includegraphics[width=0.65\textwidth]{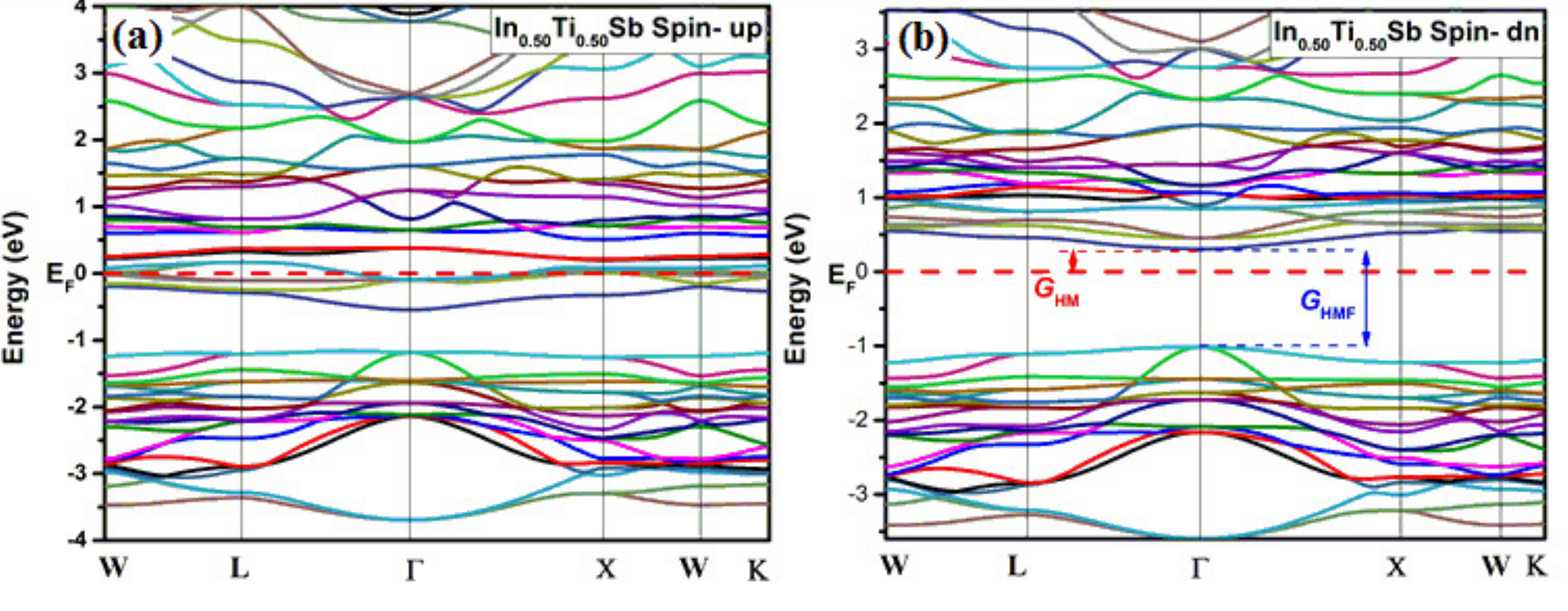}}
\caption{(Colour online) Spin-polarized band structure obtained with TB-mBJ for In$_{0.50}$Ti$_{0.50}$Sb: (a)~majority spin (up) and (b) minority spin (dn). The Fermi level is set to zero (horizontal dotted red line).} \label{fig7}
\end{figure}

\begin{figure}[htb]
\centerline{\includegraphics[width=0.65\textwidth]{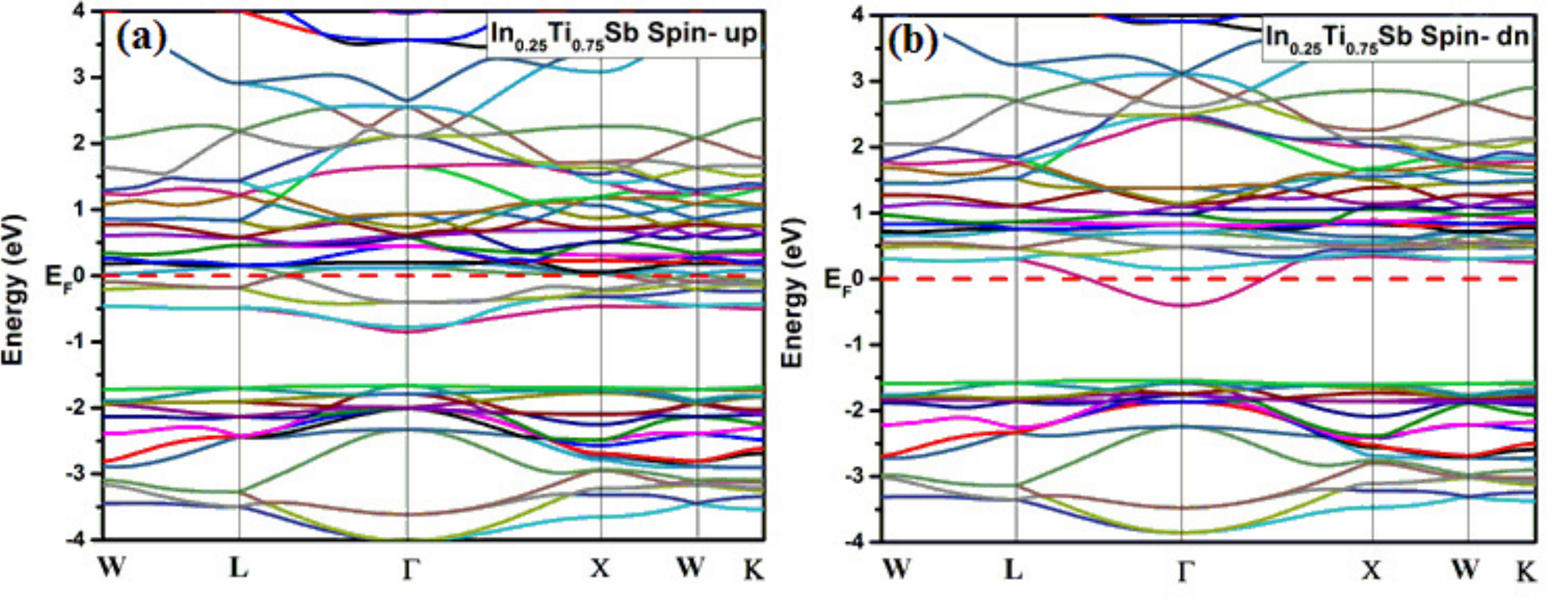}}
\caption{(Colour online) Spin-polarized band structure obtained with TB-mBJ for In$_{0.25}$Ti$_{0.75}$Sb: (a)~majority spin (up) and (b) minority spin (dn). The Fermi level is set to zero (horizontal dotted red line).} \label{fig8}
\end{figure}

\subsection{Electronic properties}
In this section, we calculated the electronic properties result in the band structure, total and partial densities of state schemes.  The calculations were performed for In$_{1-x}$Ti$_{x}$Sb with different composition $x$ ($x=0$, $0.125$, $0.25$, $0.50$, $0.75$, $0.875$ and $1$), employing GGA-PBEsol associated with TB-mBJ of Tran--Blaha modified Becke--Johnson approximation~\cite{koller2011merits,koller2012improving,tran2009accurate,gao2007half}. Figure~\ref{fig4}~(a) and figure~\ref{fig4}~(b) exhibit the spin-polarized band structure of the binary InSb ($x=0$), and display that the two spin channels of InSb have identical semiconducting band structures with a direct bandgap  located between and  high-symmetry points. The band structures of In$_{0.875}$Ti$_{0.125}$Sb,  In$_{0.75}$Ti$_{0.25}$Sb, and In$_{0.50}$Ti$_{0.50}$Sb are presented in figures~\ref{fig5}, \ref{fig6}, and \ref{fig7}. We notice that the majority-spin bands are metallic with respect to the gap located at the Fermi level for the minority-spin channel.
 \begin{figure}[htb]
 	\centerline{\includegraphics[width=0.65\textwidth]{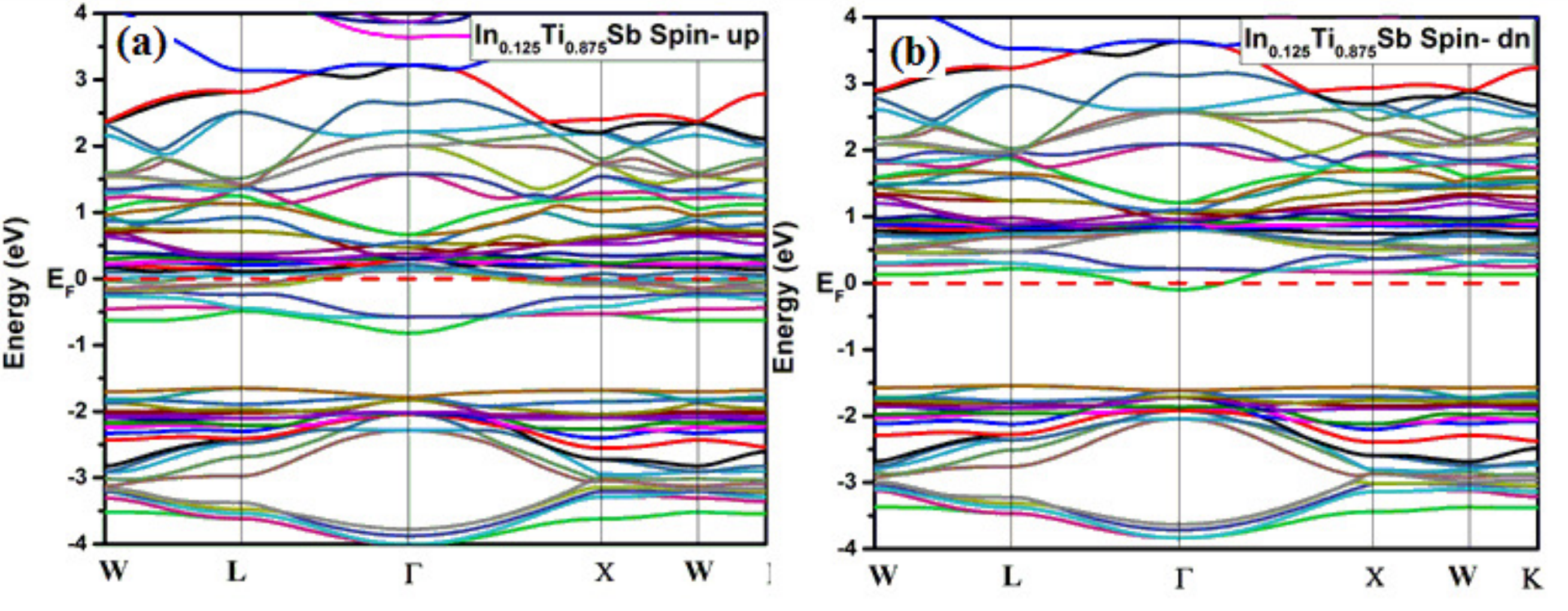}}
 	\caption{(Colour online) Spin-polarized band structure obtained with TB-mBJ for In$_{0.125}$Ti$_{0.875}$Sb: (a)~majority spin (up) and (b) minority spin (dn). The Fermi level is set to zero (horizontal dotted red line).} \label{fig9}
 \end{figure}
 
 \begin{figure}[htb]
 	\centerline{\includegraphics[width=0.65\textwidth]{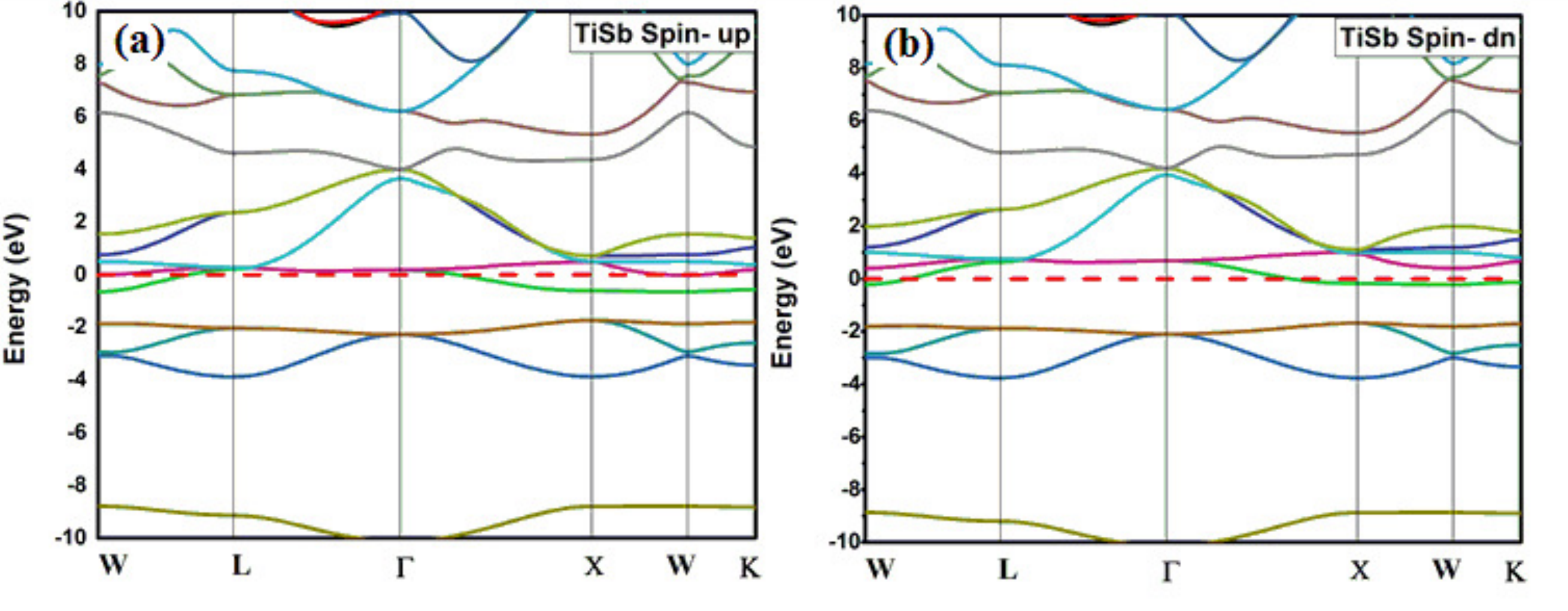}}
 	\caption{(Colour online) Spin-polarized band structure obtained with TB-mBJ for TiSb: (a) majority spin (up) and (b) minority spin (dn). The Fermi level is set to zero (horizontal dotted red line).} \label{fig10}
 \end{figure}
 
 \begin{figure}[htb]
 	\centerline{\includegraphics[width=0.35\textwidth]{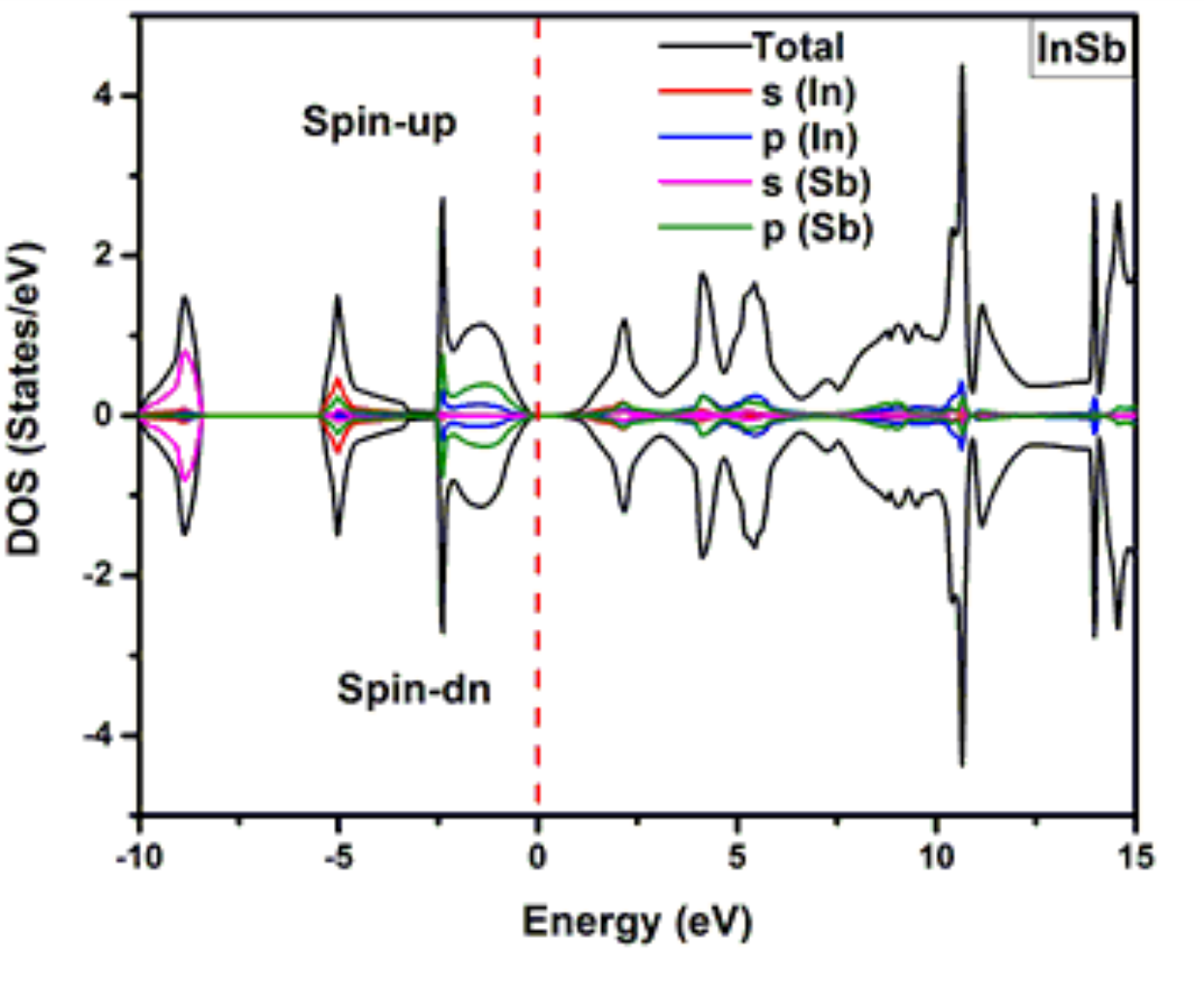}}
 	\caption{(Colour online) Spin-polarized total densities of states of InSb. The Fermi level is set to zero (vertical dotted red line).} \label{fig11}
 \end{figure}
 
 \begin{figure}[htb]
 	\centerline{\includegraphics[width=0.65\textwidth]{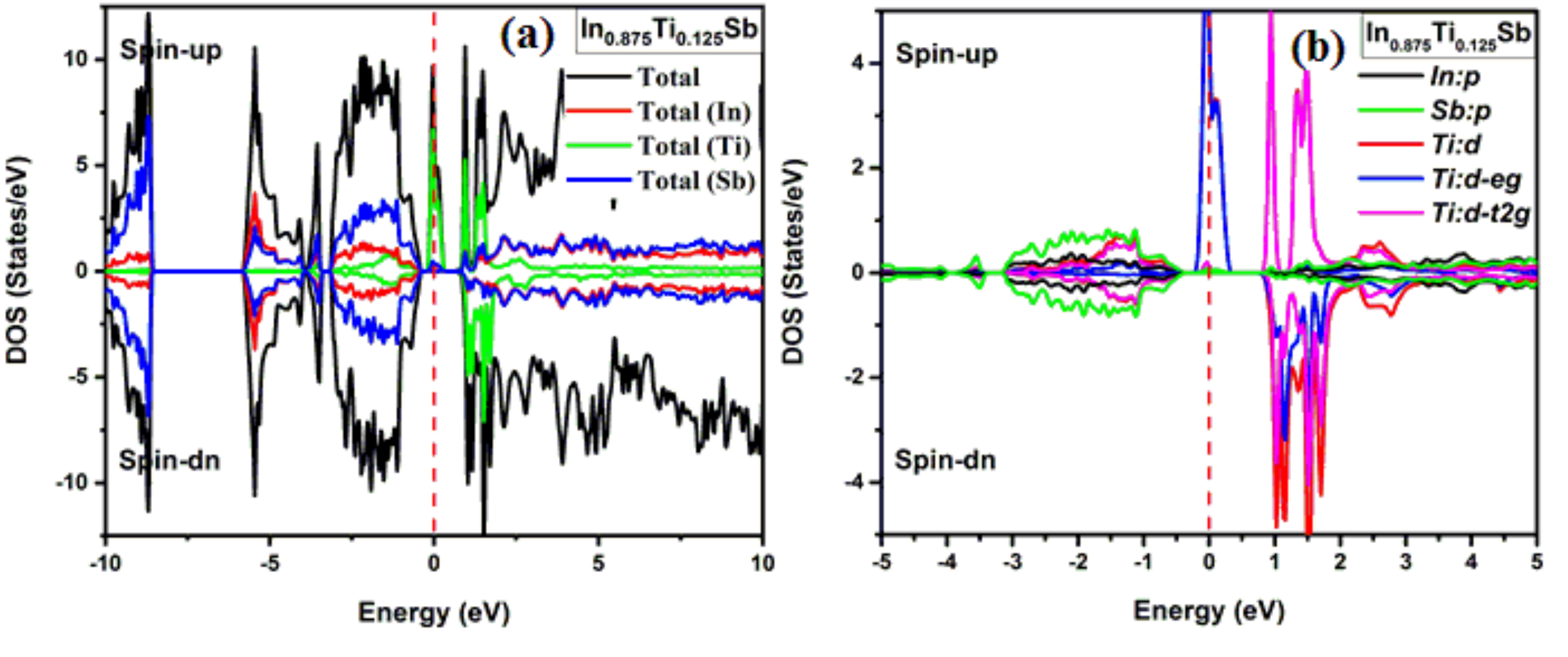}}
 	\caption{(Colour online) Spin-polarized (a) total and (b) partial densities of states of In$_{0.875}$Ti$_{0.125}$Sb.  The Fermi level is set to zero (vertical dotted red line).} \label{fig12}
 \end{figure}
 
 \begin{figure}[htb]
 	\centerline{\includegraphics[width=0.65\textwidth]{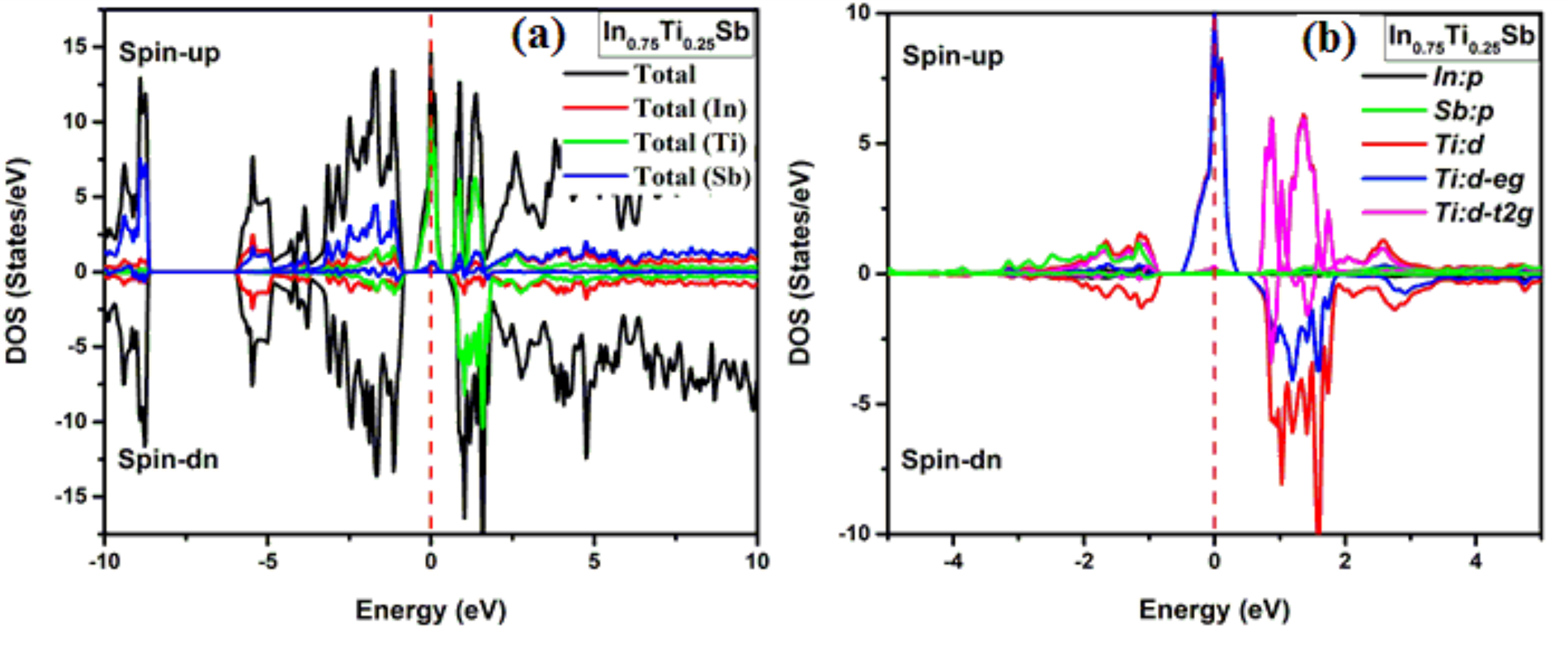}}
 	\caption{(Colour online) Spin-polarized (a) total and (b) partial densities of states of In$_{0.75}$Ti$_{0.25}$Sb.      The Fermi level is set to zero (vertical dotted red line).} \label{fig13}
 \end{figure}

\begin{table}[!htb]
\caption{The calculated indirect bandgap  for InSb, half-metallic ferromagnetic gap $G_{\rm{HMF}}$ and half-metallic gap $G_{\rm{HM}}$ of minority-spin bands for In$_{1-x}$Ti$_{x}$Sb at concentrations $x = 0$, $0.125$, $0.25$, $0.50$, $0.75$, $0.875$ and $1$.}
\label{tbl2}
\vspace{2ex}
\begin{center}
\vspace{2ex}
\begin{tabular}{|c|c|c|c|c|c|}
\hline 
\small Compound & $G_{\rm{HMF}}$ (eV) & \small $G_{\rm{HM}}$ (eV) &\small  $E^{\Gamma-\Gamma}$(eV) & \small Method & \small Behavior \\ 
\hline 
\small This work &  &  &  & \small GGA-WC &  \\ 
\hline 
\small InSb &  &  & \small 0.3216 &  &  \\ 
\hline 
\small In$_{0.875}$Ti$_{0.125}$Sb &\small  0.8423 &\small  0.3989 &  &  & \small HMF \\ 
\hline 
\small In$_{0.75}$Ti$_{0.25}$Sb & \small 1.3635 & \small 0.5182 &  &  & \small HMF \\ 
\hline 
\small In$_{0.50}$Ti$_{0.50}$Sb & \small 1.3194 & \small 0.3072 &  &  & \small HMF \\ 
\hline 
\small In$_{0.25}$Ti$_{0.75}$Sb & - & - &  &  & \small Metallic \\ 
\hline 
\small In$_{0.125}$Ti$_{0.875}$Sb  & - & - &  &  & \small Metallic \\ 
\hline 
\small TiSb & - & - &  &  & \small Metallic \\ 
\hline 
\small Other calculations &  &  &  &  &  \\ 
\hline 
\small InSb &  &  & \small 0.24~\cite{singh2017study} & \small GGA-WC &  \\ 
\hline 
 &  &  & \small 0.23, 0.2169, 0.1947, 0.17~\cite{ohno1996ga}& \small Experimental &  \\ 
\hline 
\end{tabular} 
\renewcommand{\arraystretch}{1}
\end{center}
\end{table}

\begin{figure}[!htb]
	\centerline{\includegraphics[width=0.65\textwidth]{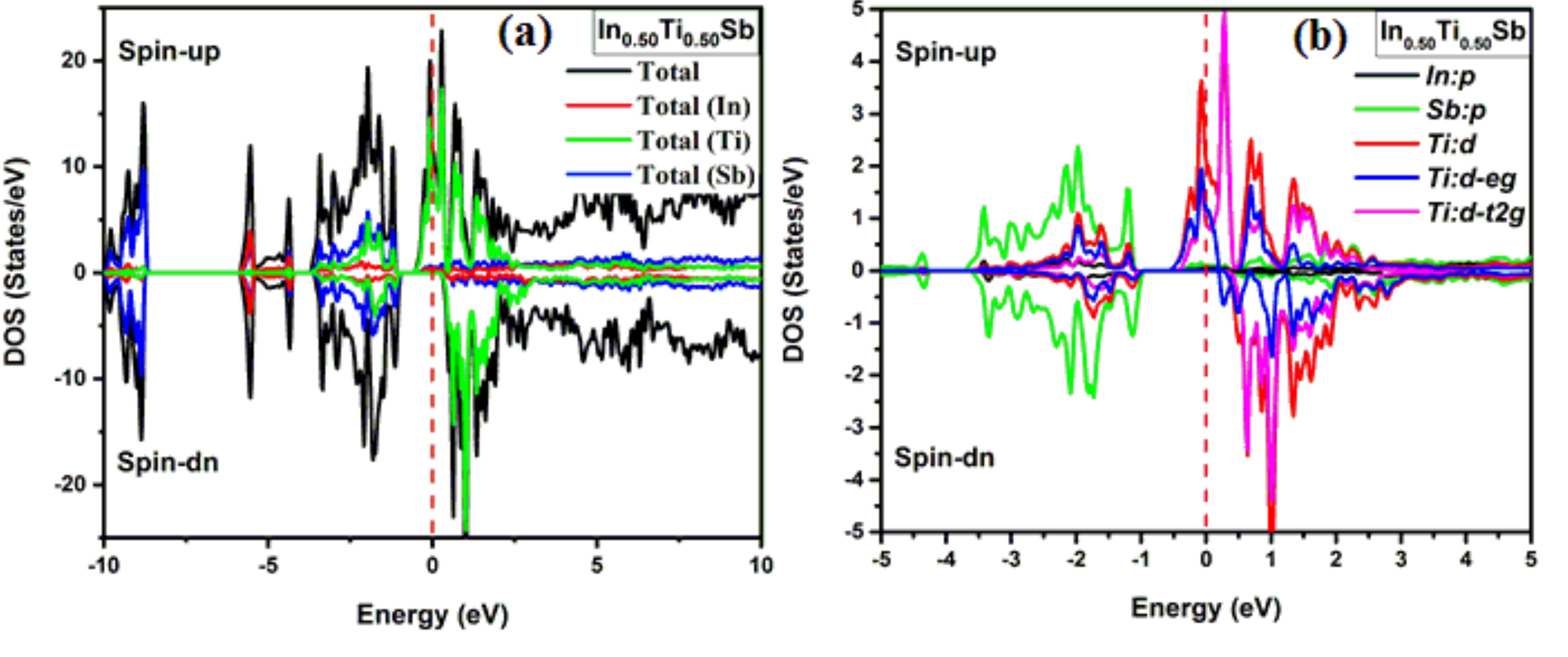}}
	\caption{(Colour online) Spin-polarized (a) total and (b) partial densities of states of In$_{0.50}$Ti$_{0.50}$Sb.   The Fermi level is set to zero (vertical dotted red line).} \label{fig14}
\end{figure}

\begin{figure}[htb]
	\centerline{\includegraphics[width=0.65\textwidth]{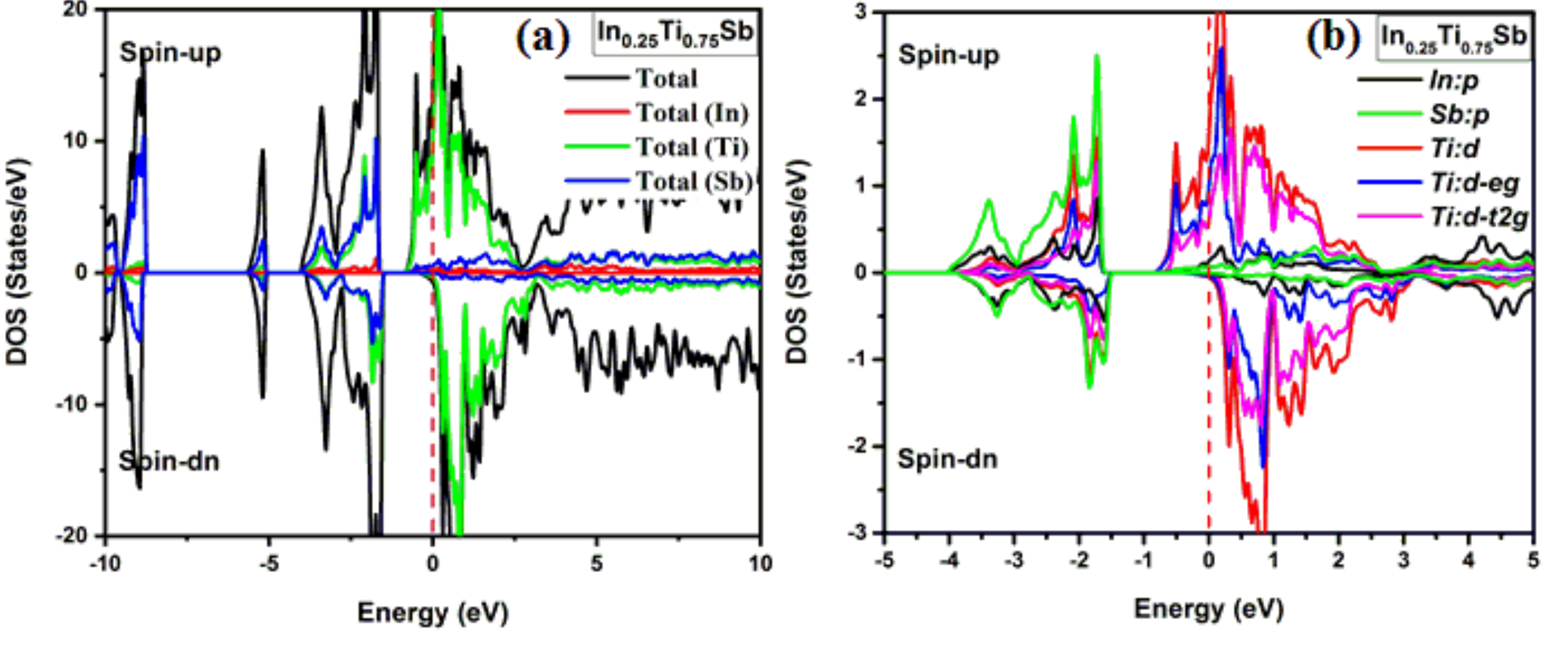}}
	\caption{(Colour online) Spin-polarized (a) total and (b) partial densities of states of In$_{0.25}$Ti$_{0.75}$Sb. The Fermi level is set to zero (vertical dotted red line).} \label{fig15}
\end{figure}

\begin{figure}[!htb]
	\centerline{\includegraphics[width=0.65\textwidth]{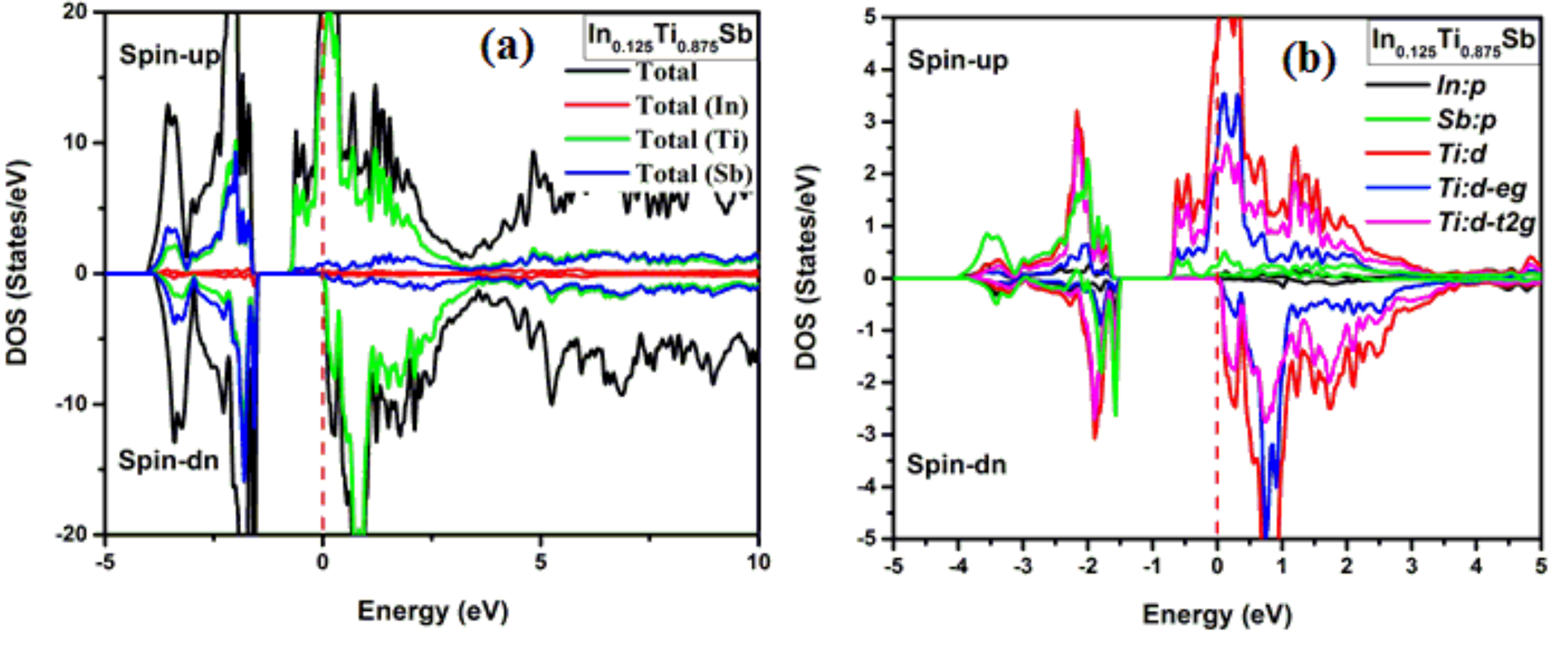}}
	\caption{(Colour online) Spin-polarized (a) total and (b) partial densities of states of In$_{0.125}$Ti$_{0.875}$Sb. The Fermi level is set to zero (vertical dotted red line).} \label{fig16}
\end{figure}

\begin{figure}[!htb]
	\centerline{\includegraphics[width=0.65\textwidth]{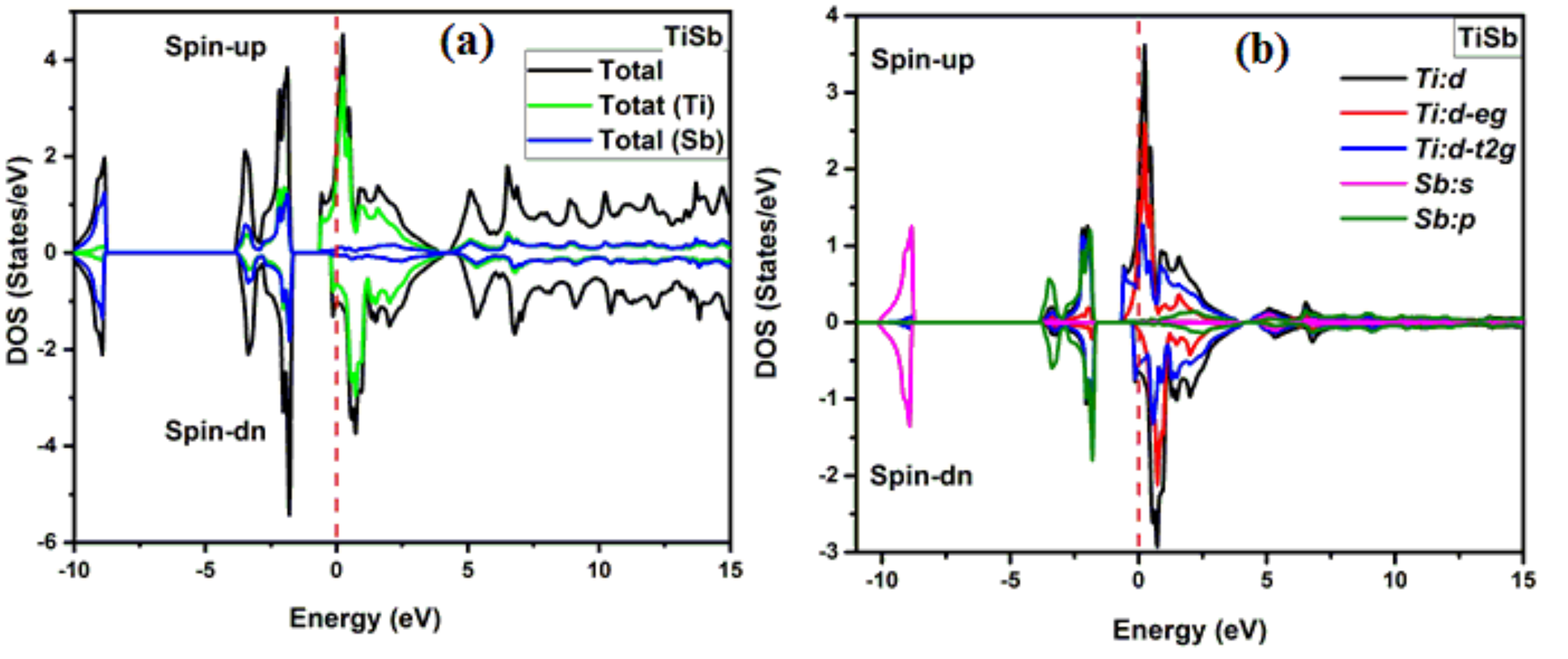}}
	\caption{(Colour online) Spin-polarized (a) total and (b) partial densities of states of TiSb. The Fermi level is set to zero (vertical dotted red line).} \label{fig17}
\end{figure}

Concerning the minority spin, we notice two gaps. The half-metallic ferromagnetic gap $G_{\rm{HMF}}$ and the half-metallic gap (i.e., spin-flip excitation) $G_{\rm{HM}}$. $G_{\rm{HMF}}$ is the energy difference between the valence-band maximum (VBM) and conduction-band minimum (CBM) while $G_{\rm{HM}}$ is the smaller energy between the minimum of the conduction band energy compared to the Fermi level and the maximum value of the valence band energy~\cite{gao2007half}. Figures~\ref{fig8}, \ref{fig9}, and \ref{fig10} shows the spin-polarized band structure of In$_{0.25}$Ti$_{0.75}$Sb, In$_{0.125}$Ti$_{0.875}$Sb, and TiSb, respectively. It is clear that the structures In$_{1-x}$Ti$_{x}$Sb for $x=0.75$, 0.875, and~1~(TiSb) have a metallic behavior for the two channels (spin up and spin down). The calculated TB-mBJ values of Eg, HMF $G_{\rm{HMF}}$, and HM $G_{\rm{HM}}$ gaps of In$_{1-x}$Ti$_{x}$Sb for different concentrations are exposed in table~\ref{tbl2} along with other experimental and theoretical data. In order to know the comportment of these alloys around the Fermi level (Fermi energy) $E_{\rm{F}}$, we have investigated the total density of states (TDOS) and partial densities of states (PDOS) of the In$_{1-x}$Ti$_{x}$Sb material. Such results are illustrated in figures~\ref{fig11}--\ref{fig17}. From these figures, it is found that binary InSb displays a semiconductor alloy, and for In$_{0.875}$Ti$_{0.125}$Sb, In$_{0.75}$Ti$_{0.25}$Sb, and In$_{0.50}$Ti$_{0.50}$Sb systems exhibit a half-metallic ferromagnetic character with spin polarization of 100\% by using the following formula given by~\cite{bai2011effects}:
\begin{align}
\label{delta-def}
P = \frac{N\uparrow(E_{\rm{F}})-N\downarrow(E_{\rm{F}})}{N\uparrow(E_{\rm{F}})+N\downarrow(E_{\rm{F}})}.  
\end{align}
Here $P$ is the spin polarization, $N\uparrow(E_{\rm{F}})$ and $N\downarrow(E_{\rm{F}})$ are the densities of states of the majority spin and minority spin around the Fermi level, respectively. In accordance with the calculated $P$ and the total DOS plot and owing to the performance that In$_{0.875}$Ti$_{0.125}$Sb, In$_{0.75}$Ti$_{0.25}$Sb, and In$_{0.50}$Ti$_{0.50}$Sb structures are equal to zero, the spin polarization is 100\%, while In$_{0.25}$Ti$_{0.75}$Sb, In$_{0.125}$Ti$_{0.875}$Sb, and TiSb demonstrates metallic comportment. It evidently appears that the titanium bands are situated at the Fermi level (plot in green). 
Figures~\ref{fig11}~(b), \ref{fig12}~(b), \ref{fig13}~(b), \ref{fig14}~(b), \ref{fig15}~(b), \ref{fig16}~(b), and \ref{fig17}~(b), depict the spin-polarized partial electronic density of states for titanium atoms in In$_{1-x}$Ti$_{x}$Sb structures. We can observe that the high section of valence bands is formed by the $3d$ (Ti) partially occupied states, and the lowest part of the conduction bands is dominated by the $3d$~(Ti) unoccupied states higher than Fermi level $E_{\rm{F}}$. 

\subsection{Magnetic and half-metallicity properties}

To explicate the specificity of the effects of the $p-d$ exchange aspect on the magnetic comportment, we calculated the total and local magnetic moments of the Ti, In, and Sb atoms as well as the interstitial sites in the In$_{1-x}$Ti$_{x}$Sb structures. Table~\ref{tbl3} exposes the total and local magnetic moments of In$_{1-x}$Ti$_{x}$Sb for different concentration $x$. We can note that the total magnetic moment for each alloy was around 1~$\mu_{\text{B}}$, usually overdue to the local magnetic moment of the Ti atom, and could be mainly attributed to the strong $p-d$ exchange interaction between $3d$ $t_{\text{2g}}$ (Ti) and $p$ (Sb) states. The calculated magnetic moment in interstitial sites in the In$_{1-x}$Ti$_{x}$Sb alloys increases with the concentration of Ti. The results also indicate that for In$_{1-x}$Ti$_{x}$Sb at all compositions except $x=0$~(i.e., InSb), the magnetic spins of the atoms between Ti and Sb and between Sb and In have opposite signs, signifying the antiparallel induced ferromagnetic interaction between Ti and Sb and between Sb and In.

\begin{table}[!htb]
	\caption{Calculated total and local magnetic moments per Ti atom of the relevant Ti, In, and Sb atoms and in the interstitial sites (in Bohr magneton $B$) for In$_{1-x}$Ti$_{x}$Sb at concentrations $x = 0$, $0.125$, $0.25$, $0.50$, $0.75$, $0.875$ and $1$.}
	\label{tbl3}
	\vspace{2ex}
	\begin{center}
		\vspace{2ex}
		\begin{tabular}{|c|c|c|c|c|c|}
			\hline 
			Compound & Total ( $\mu_{\text{B}}$) & Ti ( $\mu_{\text{B}}$)& In ( $\mu_{\text{B}}$) & Sb ( $\mu_{\text{B}}$) & Interstitial ( $\mu_{\text{B}}$) \\ 
			\hline 
			InSb & 0 & - & 0 & 0 & 0 \\ 
			\hline 
			In$_{0.875}$Ti$_{0.125}$Sb & 1 & 0.8584 & 0.0023 & $-0.0164$ & 0.1925 \\ 
			\hline 
			In$_{0.75}$Ti$_{0.25}$Sb & 1 & 0.8462 & 0.0046 & $-0.0266$ & 0.3864 \\ 
			\hline 
			In$_{0.50}$Ti$_{0.50}$Sb & 1 & 0.8641 & 0.0151 & $-0.1164$ & 0.8054 \\ 
			\hline 
			In$_{0.25}$Ti$_{0.75}$Sb & 1 & 0.8063 & 0.0343 & $-0.0635$ & 1.4459 \\ 
			\hline 
			In$_{0.125}$Ti$_{0.875}$Sb & 1 & 0.8477 & 0.0416 & $-0.1295$ & 1.6442 \\ 
			\hline 
			TiSb & 1 & 0.70123 & - & $-0.04942$ & 0.1800 \\ 
			\hline 
		\end{tabular} 
		\renewcommand{\arraystretch}{1}
	\end{center}
\end{table}

\section{Conclusion}

In this paper, we used the Full-Potential Linearized Augmented Plane Wave (FP-LAPW+lo) method to predict the structural, electronic, and magnetic properties of the substitutional doping of Titanium on the binary InSb alloy. In$_{0.875}$Ti$_{0.125}$Sb, In$_{0.75}$Ti$_{0.25}$Sb, In$_{0.50}$Ti$_{0.50}$Sb, In$_{0.25}$Ti$_{0.75}$Sb, and 
In$_{0.125}$Ti$_{0.875}$Sb structures were evaluated. The calculated formation and cohesive energy demonstrates that In$_{0.875}$Ti$_{0.125}$Sb, In$_{0.75}$Ti$_{0.25}$Sb, In$_{0.50}$Ti$_{0.50}$Sb, In$_{0.25}$Ti$_{0.75}$Sb, and 
In$_{0.125}$Ti$_{0.875}$Sb alloys are stable and can be synthesized. We used GGA-PBEsol approximation to predict the ground-state properties. The calculated structural parameters of InSb are in good agreement with the available theoretical and experimental data. GGA-PBEsol coupled with (TB-mBJ) of Tran-Blaha modified Becke-Johnson approximation was used to estimate electronic and magnetic properties. We demonstrated that In$_{0.875}$Ti$_{0.125}$Sb, In$_{0.75}$Ti$_{0.25}$Sb, and  In$_{0.50}$Ti$_{0.50}$Sb are half-metallic ferromagnetic structures with a spin polarization of 100\% at the Fermi level, while In$_{0.25}$Ti$_{0.75}$Sb, and 
In$_{0.125}$Ti$_{0.875}$Sb present a metallic comportment. The total magnetic moments are 1~$\mu_{\text{B}}$ for all compounds. The investigation of magnetic properties demonstrates that the $p-d$ exchange systems operate in stabilizing ferromagnetic state configuration, and are the origin of a low magnetic moment in the investigated alloys. According to our bibliographic research, there are no earlier theoretical or experimental studies on the In$_{1-x}$Ti$_{x}$Sb materials. Thus, we hope that our results serve as a reference for future theoretical and experimental research.


	\ukrainianpart

	\title{Дослідження структурних, магнітних і електронних властивостей та напівметалічності твердих розчинів In$_{1-x}$Ti$_{x}$Sb на основі методу функціоналу густини}%

	\author{С. Амрані\orcid{0000-0002-3484-6480}\refaddr{label1}, М. Бербер\orcid{0000-0003-1285-3070}\refaddr{label2,label3}, М. Мебрек\orcid{0000-0002-4116-1332}\refaddr{label2,label3}}
	\addresses{
		\addr{label1} Лабораторія фізико-хімічних досліджень, Університет Саїди, 20000 Саїда, Алжир
		\addr{label2} Лабораторія перспективних матеріалів і приладобудування, Університетський центр Нур Ель-Башир Ель-Баядх, Алжир
		\addr{label3} Університетський центр Нур Ель-Башир Ель-Баядх, Ель-Баядх, 32000, Алжир
	}
	%
	
\newpage
\makeukrtitle
		
		\begin{abstract}
			
			З метою з'ясування впливу заміщення в легованому атомами Ti сплаві InSb проведено першо\-прин\-цип\-ні розрахунки в рамках методу лінеаризованих розширених плоских хвиль з повним потенціалом (ЛРПХзПП). З використанням узагальненого градієнтного наближення Пердью-Бурке-Ернцергофа для твердих тіл (УГН-ПБЕтт)			
			в покращеному підході Тран-Блаха з модифікацією Бека-Джонсона (ТБ-мБД)
			передбачено структурні, електронні і магнітні властивості сполук  In$_{1-x}$Ti$_{x}$Sb з концентраціями $x = 0$ , $0.125$, $0.25$, $0.50$, $0.75$, $0.875$ та $1$. Розраховані значення сталих ґратки добре узгоджуються з наявними теоретичними та експериментальними даними. Як показують розрахунки, усі структури є енергетично стійкими. Легування заміщення  перетворює іонний характер вихідної сполуки InSb на напівметалічну феромагнітну поведінку для концентрацій $x = 0$, $0.125$, $0.25$ і $0.50$, зі $100$\% спіновою поляризацією на рівні Фермі та на металічну для In$_{0.25}$Ti$_{0.75}$Sb і  In$_{0.125}$Ti$_{0.875}$Sb. Оціночні значення повних магнітних моментів становлять приблизно 1~$\mu_{\text{B}}$.  $In_{0.875}$Ti$_{0.125}$Sb,  In$_{0.75}$Ti$_{0.25}$Sb, та  In$_{0.50}$Ti$_{0.50}$Sb демонструють напівметалічну феромагнітну поведінку та потенційно можуть використовуватися в спінтроніці.

\end{abstract}

\lastpage
\end{document}